\documentstyle[12pt,epsfig]{article}
\def\rsim{\mathrel{\raise2pt\hbox to 8pt{\raise -6pt\hbox{$\sim$}\hss{$>$}}}}
\def\lsim{\mathrel{\raise2pt\hbox to 8pt{\raise -6pt\hbox{$\sim$}\hss{$<$}}}}

\begin{document}

\begin{center}
{\Large{\bf Chiral Symmetry in Nuclei \rm}\rm} \\

\vspace*{0.25in} 

{\bf J.\ L.\ FRIAR \rm} \\
{\it Theoretical Division \\
Los Alamos National Laboratory \\
Los Alamos, NM  87545 USA}

\end{center}
\begin{abstract}

The impact of chiral symmetry on nuclear physics is discussed in the context of
recent advances in the few-nucleon systems and of dimensional power counting.
The tractability of few-nucleon calculations, illustrated by very recent
solutions for $A = 2-6$, is shown to follow from power counting based on chiral
Lagrangians. The latter predicts the suppression of $N$-body forces, as
originally shown by Weinberg.  Isospin violation in the nuclear force is
similarly analyzed using the results of van Kolck, and this is shown to be
consistent with results from the Nijmegen phase-shift analysis.  Conventional
$\rho - \omega$ and $\pi - \eta$ mixing models with on-shell mixing strength are
not inconsistent with naive power counting. Meson-exchange currents calculated
in chiral perturbation theory are in good agreement with experiment.

\end{abstract}

\section{Introduction}

My talk will try to merge two rather different areas of physics:  classical
(conventional) nuclear physics and modern field-theory-based particle physics.
This is not an easy task.  A skeptic might say that if you wanted to invent a
system with little \underline{apparent} dynamical basis, great complexity of
internal structure (spin and isospin), and almost pathological computational
difficulty, you would call it a nucleus!  This view was purposefully overstated,
but has elements of truth nevertheless.  Nuclear dynamics was mostly
phenomenological for decades.  The dominance of the tensor force and the
one-pion-exchange potential (OPEP) means that spin and isospin play an essential
role at leading order and intuition based on simple central forces doesn't often
apply.  This internal structure makes computational problems exceptionally
challenging, which has inhibited our ability to solve (numerically) the
Schr\"odinger equation. The net result was that in order to learn about
strong-interaction dynamics in a nucleus we had to be able to calculate, and we
couldn't do the latter with much accuracy. 

Recent computational advances\cite{1} in treating few-nucleon systems may have
broken this logjam.  In spite of all the difficulties we have finally begun to
realize the potential inherent in studying few-nucleon systems.  This area of
physics on which I will focus is in my opinion the biggest success story in
nuclear physics in the past decade.  Beginning about ten years ago, we have made
spectacular progress in solving (numerically) most of the seminal problems that
were discussed decades ago as crucial to the success of the field.  New
terminology was coined, with ``exact'' or ``complete'' denoting calculations of
observables with errors of less than 1\% (in spite of the computational
hardships). Problems are now being solved that were considered far out of our
reach ten years ago.  This work is beginning to yield dynamical information,
which will undoubtedly pay dividends in the future.  In order to illustrate how
things have changed, almost everything that we have calculated ``works'', with
those few disagreements with experiment being closely examined and debated and
providing considerable hope for more progress. 

My purview is chiral symmetry (CS) in nuclear physics.  Others much more
knowledgeable than I am have talked about the particle physics aspects,
including the fashionable and successful Chiral Perturbation Theory ($\chi$PT).
I hope to be able to convince you that this symmetry has a dominant influence in
nuclear physics.  Without the symmetry, nuclear physics would be intractable.
Indeed, we can and will turn this argument around: the tractability of nuclear
physics provides a strong signature for the effect of CS in nuclei. Chiral
perturbation theory could turn out to be the biggest advance in nuclear physics
in decades, or of very limited use.  There is a huge amount of information
contained in our field on the behavior of the strong interactions, and we have a
great opportunity for unifying all of hadronic physics.  The theoretical
approach used in nuclear physics unfortunately lacks (in part) the well-defined
methodology of particle physics, and the challenge will be to try to change
this. 

I can summarize the talk by stating that CS and dimensional power counting have
an opinion about: (1) the sizes of various components of the nuclear force (both
isospin-conserving and isospin-violating); (2) the relative size of
three-nucleon forces(3Nf); (3) the relative size of four-nucleon forces(4Nf),
... ; (4) the relative size of relativistic corrections in (light) nuclei; (5)
the relative size of nucleon (impulse approximation) and meson-exchange currents
in nuclear electromagnetic and weak interactions. 

\section{Nuclear Physics Overview}

Any discussion of the role of chiral symmetry in nuclear physics must begin with
a brief discussion of three topics that will tell us how nuclear theorists do
business and possibly how this should change:  what we do, why we do it, and
what we need to do.  For the purposes of this talk when I say ``nuclear
physics'', I mean ``low-energy few-nucleon physics'', unless stated otherwise.
Restricting myself to the traditional domain of nuclear physics ($\lsim$ a few
hundreds of MeV) frames the problem sufficiently for my allotted time. 

Potentials used in the context of the Schr\"{o}dinger equation (or a
generalization) are central to the organization of nuclear calculations.
Dynamics, assumptions, and prejudices are contained in this quantity.  The
reason why potentials are used is twofold and simple:  (1) nuclei are
self-bound configurations of nucleons, and binding cannot be achieved in
(finite-order) perturbation theory; (2)  when two nucleons interact and
propagate between interactions there is an infrared singularity that enhances
successive iterations, and this is treated exactly by the Schr\"{o}dinger
equation\cite{2}. Thus our scheme is extremely efficient and reduces the
complexity of calculating an amplitude to that of defining a potential. 

Although the underlying dynamics of nuclei and elementary particles is shared,
the two are rather different in their scales.  Simply stated, nuclei are large,
squishy, and soft, while particles are small, stiff, and hard.  These are words
used to state that the radii of nuclei follow $R \simeq 1.2A^{\frac{1}{3}}$ fm,
while particles are smaller than 1 fm.  The excitation energies of nuclei are
typically tens of MeV or less, while particles require hundreds of MeV, and
nuclear internal momenta are \underline{on average} fairly small, while in
particles they can be high.  We can estimate the latter, $\overline{p}$, using
the uncertainty principle in the He isotopes\cite{3}. Equating $\overline{p} R
\sim \hbar$ and $R \sim$ 1.5-2.0 fm, we obtain $\overline{p} c \sim$ 100-150
MeV. For mnemonic purposes \underline{only}, one can equate this to the pion
mass: $\overline{p}c \sim m_{\pi} c^2$. This is clearly inappropriate in the
chiral limit, $m_{\pi} \rightarrow 0$. Note that this value is about half of the
Fermi momentum, $\hbar k_F c =$ 260 MeV, which characterizes nuclear matter.
Momentum components larger than this can play a significant role in some cases
and the estimate should not be taken too literally. 

Given this scale for momenta there are other scales that can be constructed.
Nucleons with mass $M$ are heavy and slow moving.  The average kinetic energy of
a nucleon is roughly $\frac{\overline{p}^{2}}{M}$ or $\sim m^2_{\pi}/M \sim$ 20
MeV, which is fairly accurate for $^2$H, $^3$H, $^3$He, and $^4$He.  Because
nuclei are weakly bound systems, potential and kinetic energies are comparable
in magnitude.  Semirelativistic calculations\cite{4} for these nuclei (using
$\sqrt{p^2 + m^2} - m$) find corrections of $\sim$ 5\% to the kinetic energy,
which are typically balanced by changes in the potential.  The dominant physics
is nonrelativistic. 

Given these scales we can easily estimate what happens when two nucleons
propagate between interactions.  The Green's function, $G$, schematically is
$1/(\overline{p}^2/M) \sim \frac{M}{m^{2}_{\pi}}$ and becomes very large for
small $\overline{p}$.  It is worth remembering that potentials (unlike
amplitudes) are not uniquely defined\cite{2}. Rather, potentials are (nonunique)
subamplitudes, and this leads to the ``off-shell'' problem of nuclear physics.
Although it is possible to set criteria for how one defines $V$, the fact that
$G^{-1}$ is small means that rather small changes in $V$ can be compensated by
the infrared singularity in $G$, leading to an alternative (definition of) $V$,
which may differ substantially. 

\begin{figure}[htb]
  \epsfig{file=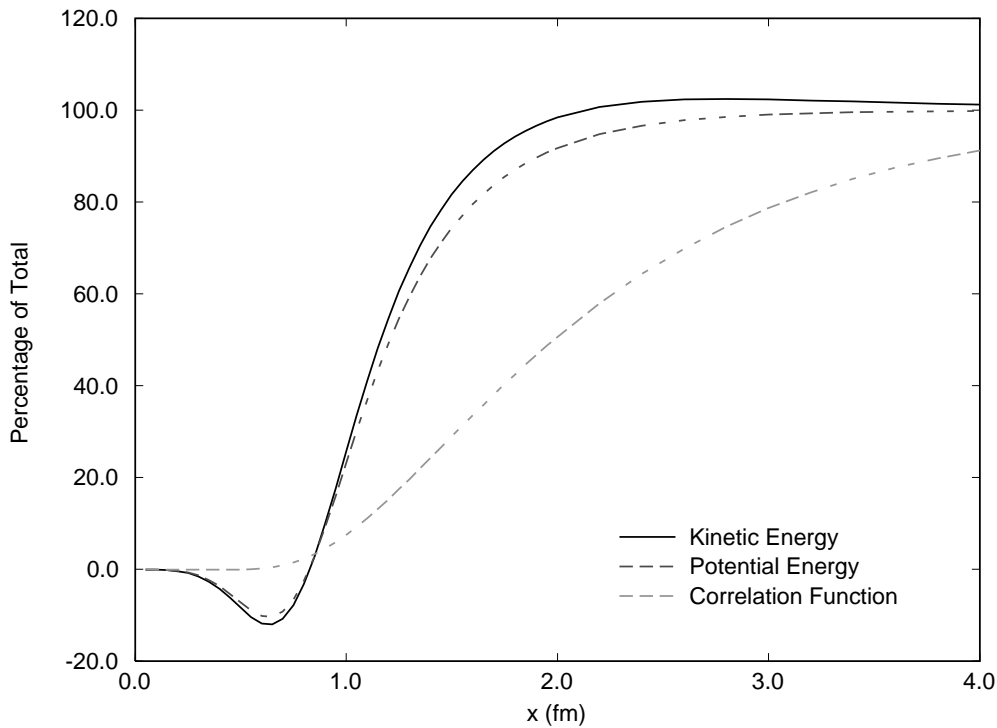,height=4.0in,bbllx=17pt,bblly=406pt,bburx=523pt,bbury=730pt}
  \caption{Percentages of accrual of kinetic energy (solid line),
   potential energy (short dashed line), and probability (long dashed line)
   within an interparticle separation, $x$, for any pair of nucleons in the
   triton. }
\end{figure}

The biggest conceptual problem in nuclear physics calculations (in my opinion)
is the lack of a well-defined regularization scheme.  The successive iteration
of two potentials (or the sequential exchange of two mesons) is given by a loop
integral, which is almost always divergent.  In order to regularize this
divergence nuclear potentials are cut off at short distances (large momenta),
which leads to short-range repulsion and renders the calculations finite.  These
cutoffs are typically for momenta $\sim$ 1 GeV, are assumed to derive from meson
clouds around the nucleons (i.e., form factors), and are treated as parameters. 
This procedure is very efficient, physically motivated, and not very well
defined, but it is the best we can do at the moment.  Because of this problem
very short-range operators that arise in \underline{ab} \underline{initio}
calculations of potentials or transition operators are sensitive to details of
the regularization procedure.  Zero-range operators (which clearly require
regularization) are either assumed not to contribute because of the short-range
repulsion (which makes a ``hole'' in the wave function), or are given a finite
size.  The results will be different. 

Figure 1 shows the result of integrating the separation of any two nucleons in 
the triton out to a distance, x, and then integrating over the coordinates of
the third nucleon.  The three curves show the kinetic energy, potential energy,
and probability (correlation function) calculated in this way, which must
approach 100\% as x increases.  The effect of short-range repulsion and the
volume element ($d^3x$) are obvious at small separation.  Fortunately, most of
the energy accrual occurs between 1 and 2 fm, corresponding to fairly small
(virtual) momenta.  Nevertheless, there are practical and conceptual problems at
short distances in all conventional treatments.  We obviously need a better way
to handle this difficulty. Although the regularization question clouds the issue
of testing chiral symmetry, it doesn't change our conclusions. 

\section{Few-Nucleon Systems}

There exists a class of recently developed nucleon-nucleon ($NN$) potentials
that fit the available $NN$ scattering data remarkably well and, in addition,
contain much important physics\cite{5,6}.  Many of the older potentials had a
number of annoying minor defects that have been removed.  These defects were
highly distracting, but probably not very important in most calculations.  One
of these new potentials, the Argonne $V_{18}$\cite{6}, has 18 well-defined
spin-isospin-orbital operators, illustrating the point about complexity that was
made in the introduction. This potential was used to calculate the ground-state
properties of the deuteron $[^2{\rm H} (1^+)]$, the triton $[^3{\rm
H}(\frac{1}{2}^+)]$, $^3{\rm He}(\frac{1}{2}^+)$, the $\alpha$-particle $[^4{\rm
He}(0^+)]$, $^5{\rm He}(\frac{3}{2}^-)$, $^5{\rm He}(\frac{1}{2}^-)$, and
$^6{\rm Li}(1^+)$, as well as the $3^+$ excited state of the latter and the
$(^6{\rm He}(0^+)$, $^6{\rm Li}(0^+)$, $^6{\rm Be}(0^+)$) isospin triplet.
Reference (7) finds the results listed in Table 1.\\ 

\begin{table}[hbt]
\centering
\caption{Calculated and experimental ground-state energies of few-nucleon 
    systems, together with (approximate) dates when they were first accurately 
    solved for ``realistic'' potentials.}

\begin{tabular}{|l||llllll|}
\hline
{Nucleus($J^{\pi})$}&\hspace{0.1in} {$^2{\rm H}(1^+)$} & 
{$^3{\rm H}(\frac{1}{2}^+)$} & 
{$^4{\rm He}(0^+)$} &  {$^5{\rm He}(\frac{3}{2}^-)$} & 
{$^5{\rm He}(\frac{1}{2}^-)$} & {$^6{\rm Li}(1^+)$} \\ \hline \hline
{First Solved} &\hspace{0.1in}$\sim$1950 & 1984 & 1987 & 1990 &
 1990 & 1994 \\ \hline
Expt. (MeV) &\hspace{0.1in} -2.22 & -8.48 & -28.3 & -27.2 & -25.8 & -32.0 \\
Theory (MeV) &\hspace{0.1in} -2.22 & -8.47(2) & -28.3(1) & -26.5(2) & -25.7(2)
& -32.4(9) \\ \hline
\end{tabular}

\end{table}
A weak three-nucleon force was added to the Hamiltonian and was adjusted to fit
the binding energy of $^3$H, just as the $NN$ force fits the $^2$H binding
energy.  The rest of the theoretical results are predictions, and are in
excellent agreement with experiment. We have also indicated when the
Schr\"odinger equation was first solved for each case. Much of the indicated
progress is recent. 

Examining the 2-, 3-, and 4-body cases, one finds that roughly 20 MeV/pair of
nucleons accrues from the $NN$ force, while approximately 1 MeV/triplet results
from the 3Nf.  If the error bar on the $^4$He result is taken as an upper limit,
the effect of any 4Nf should be less than .1 MeV/quartet.  These numbers will be
interpreted in Section 7. As spectacular as the results are, even better things
are being planned. The A = 7 and 8 systems should be tractable when improved
computers become available in the near future. These calculations demonstrate 
the recent achievements in the few-nucleon field. 

\section{Pion Degrees of Freedom}

For much of its existence nuclear physics made the tacit assumption that
nucleons are the only significant degrees of freedom manifested in nuclei.  The
reason is simple:  this paradigm works, and works well.  In order to demonstrate
conclusively the contribution of other degrees of freedom, one must show that
trustworthy calculations fail to reproduce experimental data. It was found long
ago that compared to the best theoretical calculations\cite{8} the experimental
cross section for the radiative capture of thermal neutrons by protons was too
large by approximately 10\%. That is, calculations of this M1 reaction were too
small if one assumed that the final photon was emitted solely from the nucleons.
Moreover, since this impulse approximation is easy to evaluate, either other
processes contribute or our understanding of the deuteron is seriously flawed.
Since the late 1940s people had suspected that mesons were involved, but no
compelling case for this scenario was made.  An influential paper by Chemtob and
Rho\cite{9} showed the importance of soft-pion theorems for resolving problems
in our understanding of strong-interaction dynamics.  Soon thereafter Riska and
Brown\cite{10} provided a compelling argument based on gauge invariance and
credible phenomenology that pions were the needed ingredient for an
understanding of the process.  Not long thereafter (and continuing until the
present time), magnetic electron scattering provided the unassailable graphic
evidence for meson currents. 

\begin{figure}[htb]
  \epsfig{file=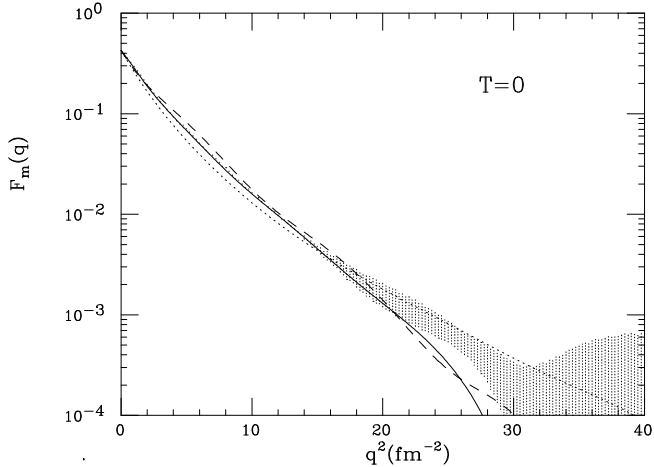,height=4.0in,bbllx=72pt,bblly=36pt,bburx=540pt,bbury=756pt,angle=90}
  \caption{Isoscalar elastic magnetic form factor of the trinucleons as a
    function of squared momentum transfer, $q^2$.  The shaded area is the data,
    the dashed line is the impulse approximation (nucleons only), while the 
    solid and dotted lines contain meson currents.}
\end{figure}

This is illustrated in Figs. (2) and (3)\cite{11}.  Figure (2) shows the
isoscalar combination of magnetic form factors (essentially their sum) of $^3$He
and $^3$H from a recent experimental analysis.  The dashed line is an
impulse approximation (nucleons only) calculation and agrees very well with the
data (shaded curve).  Uncharged mesons (corresponding to T=0) apparently don't
play a large role.  In contradistinction, the isovector combination of form
factors (essentially their difference) in Fig. (3b) shows an impulse
approximation calculation shifted far from the data.  Inclusion of meson
currents (which are dominated by single-pion exchange) corrects the problem. 
Figure (3a) shows the threshold (transition) magnetic form factor of the
deuteron. This $^3{\rm S}_1 - ^3\!{\rm D}_1 \rightarrow ^1\!\!{\rm S}_0$
reaction is just the inverse of the thermal $np$ radiative capture.  A similar
pattern is found, which demonstrates conclusively the existence of pion degrees
of freedom in the nucleus interacting with the external fields. 

\begin{figure}[htb]
  \epsfig{file=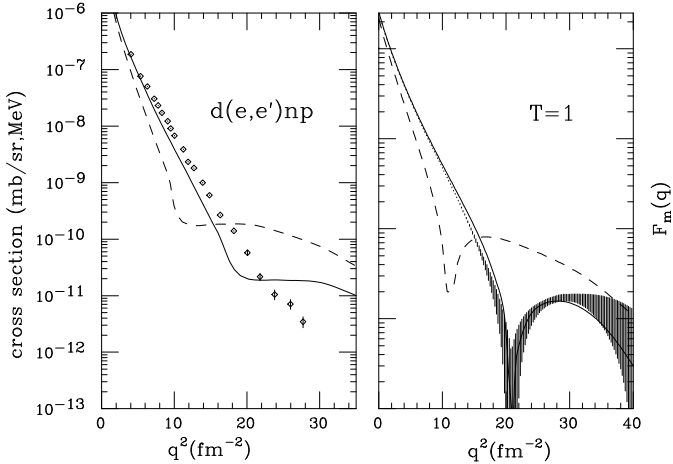,height=4.0in,bbllx=72pt,bblly=54pt,bburx=540pt,bbury=786pt,angle=90}
  \caption{The (isovector) threshold deuteron electrodisintegration is shown
in (a) as a function of squared momentum transfer, $q^2$.  The individual points
are the data, the dashed line is the impulse approximation (nucleons only),
while the solid line contains pion-exchange currents. The isovector elastic
magnetic form factor of the trinucleons is shown in (b) as a function of squared
momentum transfer, $q^2$.  The shaded area is the data, the dashed line is the
impulse approximation (nucleons only), while the solid line contains
pion-exchange currents.}
\end{figure}

That $\Delta T = 1$ reactions should display the effect of intranuclear motion
of charged mesons is perhaps no surprise, since the current continuity equation
$$ \mbox{\boldmath $\nabla$} \cdot \mbox{\boldmath $J$}_{MEC} (\mbox{\boldmath
$x$})  = -i [V, \rho(\mbox{\boldmath $x$})] \, ,\eqno (4.1) 
$$ 
relates the meson-exchange currents (MEC) to noncommuting parts of the potential
(mostly isospin dependent). That ``pure'' one-pion exchange completely dominates
is more surprising.  This was noted by Rho\cite{12}, who called the phenomenon a
``chiral filter''.  Recently, a prescription was developed\cite{13} to enforce
Eq.(4.1).  It can be shown that in the general case this prescription produces a
nearly pointlike $\pi NN$ vertex, in keeping with the ``chiral filter'', and
works quite well in most cases. We will discuss meson-exchange currents 
further in Section 7. 

The final unambiguous demonstration of pion degrees of freedom in nuclei has an
added cachet: it comes with error bars.  Beginning approximately fifteen years
ago the Nijmegen group\cite{14} have implemented a sophisticated and successful
program of Phase Shift Analysis (PSA) of the $NN$ interactions. Their
methodology includes treating all known long-range components of the
electromagnetic interaction, such as Coulomb, magnetic moment, vacuum
polarization, etc., as well as the tail of the $NN$ interaction beyond 1.4 fm,
which includes OPEP.  The inner interaction region is treated in a
phenomenological fashion.  This allows an accurate determination of the $\pi NN$
coupling constants.  In order to check for systematic errors they also fit the
masses of the exchanged pions, both charged and neutral, and find 
$$
m_{\pi^{\pm}} = 139.4 (10) \, {\rm MeV} \, , \eqno (4.2a)
$$
$$
m_{\pi^{0}} = 135.6 (13) \, {\rm MeV} \, . \eqno (4.2b)
$$
The small error bars ($\lsim$ 1\%) demonstrate the importance of OPEP in the 
nuclear force.  They are currently investigating the tail of the rest of the 
$NN$ interaction. 

A valuable byproduct of this work is the ability to construct potentials by
directly fitting to the data, rather than to phase shifts, and to utilize the
entire $NN$ data base.  Several potential models, such as the Argonne $V_{18}$
model, have been constructed in this way and fit the $NN$ data base far
better than any previous attempts.  One useful corollary of this work\cite{5} is
that a baseline has been set for the triton binding energy ($\sim$ 7.62 MeV)
using local $NN$ potentials.  Nonlocal potential components arising from
relativity are currently under intensive investigation. 

Finally, we should ask what the other consequences might be of this great
sensitivity to OPE processes.  In 1984 it was noted\cite{15} that a ``pure''
OPEP used in certain deuteron reactions was as good as using a ``realistic''
potential.  In the triton this force was substituted for the $^3{\rm S}_1 -
^3\!{\rm D}_1$ part of the potential and produced nearly the same binding as a
realistic potential.  Because that partial wave accounts for $\sim \frac{3}{4}$
of the triton potential energy, it was deduced that OPEP dominates the triton
binding. This has been subsequently quantified and extended to other
systems\cite{16}. One finds that $\langle V_{\pi} \rangle /\langle V \rangle
\sim$ 70-80\% for a wide variety of calculations ranging from ``exact''
treatments of the triton and $\alpha$-particle to variational treatments of
nuclear matter. Pion exchange is clearly of exceptional importance in nuclei,
largely due to its rather long range (it is a pseudo-Goldstone boson) and spin
($0^-$), which produces a tensor force in leading order. 

\section{Nuclear and Chiral Scales and Interactions}

We have already argued that the \underline{average} nuclear momentum scale is
$\overline{p} c \sim m_{\pi} c^2$.  Several other scales are important.  The
pion mass sets the scale for chiral-symmetry breaking.  The pion decay constant,
$f_{\pi}$ = 92.4 MeV, sets the scale for pion interactions.  The large-mass
scale, $\Lambda \sim$ 1 GeV, comes from several different sources:  the nucleon
mass $M$, the masses of all heavy mesons (or resonances) such as $\rho$ and
$\omega$, and $4 \pi f_{\pi}$, which arises naturally in loop integrals.  In any
(low-energy) process constrained by chiral symmetry, we expect that the
dimensionless parameter (``small momentum''/$\Lambda$) controls the physics and,
indeed, the convergence of any power-series expansion\cite{17}. 

In 1990 Weinberg\cite{18} introduced chiral perturbation theory into nuclear
physics by calculating the leading-order $NN$ force and the leading-order $3N$
force.  Perhaps more important was the dimensional-power-counting scheme that he
introduced and refined.  The latter is a powerful tool that categorizes
amplitudes (or potentials) in terms of powers of their characteristic energy or
momentum scales: [``small momentum''/$\Lambda$(large-mass scale)], as we noted 
above.

\begin{figure}[htb]
  \epsfig{file=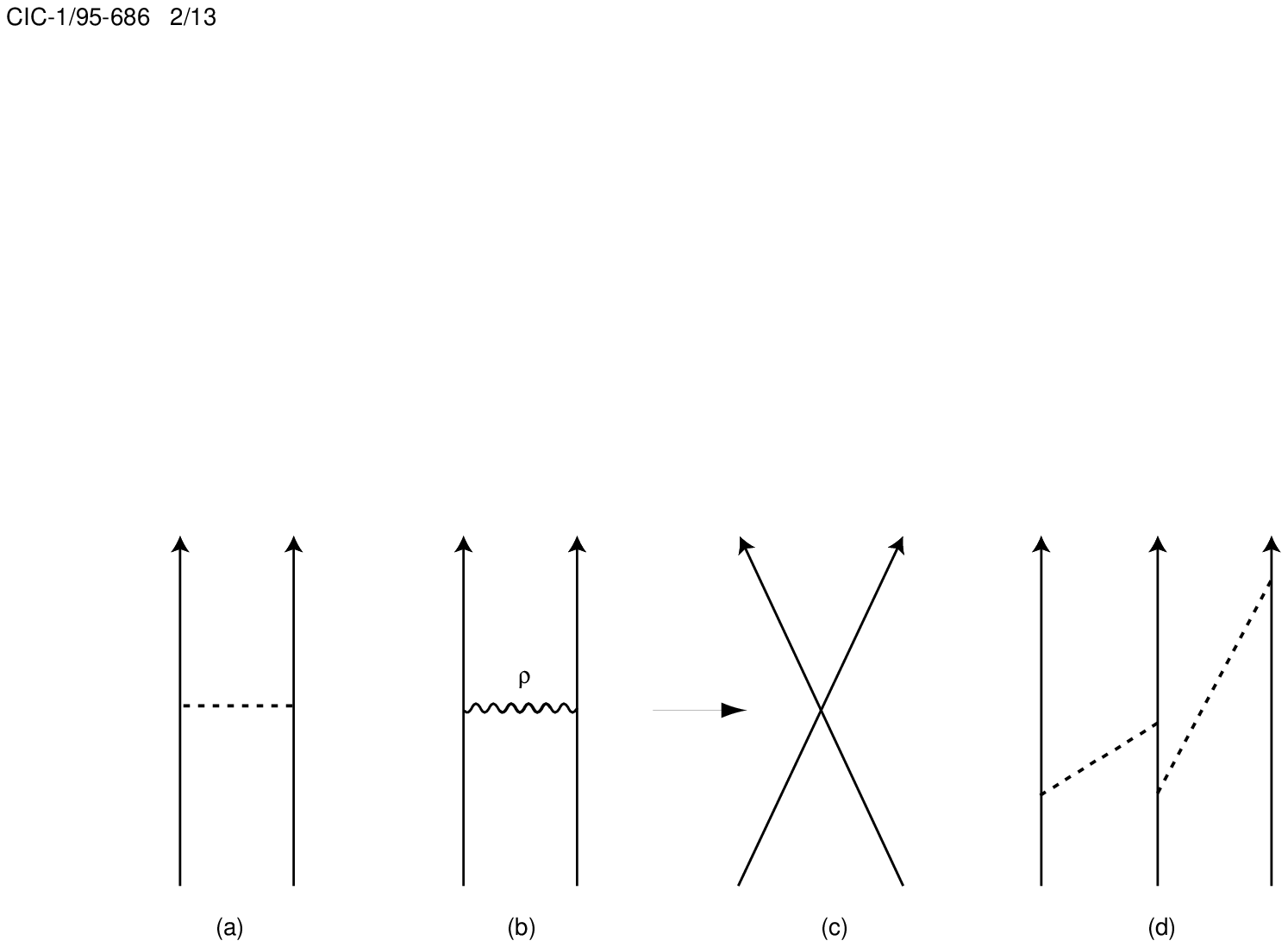,height=2.0in,bbllx=41pt,bblly=435pt,bburx=483pt,bbury=600pt,clip=}
  \caption{Time-ordered perturbation theory diagrams for nuclear potentials
in $\chi$PT, with OPEP shown in (a), $\rho$-exchange in (b) becomes a contact
interaction in (c), while overlapping pion exchanges contribute to the 3Nf in
(d). Pions are depicted by dashed lines, while nucleons are shown as solid 
lines.}
\end{figure}

The leading-order nucleon-nucleon potentials are the august one-pion-exchange
potential shown in Fig.(4a) and a generic short-range interaction for each spin
and isospin channel (shown in Fig.(4c)), which subsumes the effect of all
short-range resonant and non-resonant meson exchanges (e.g., Fig.(4b)). The
leading-order three-nucleon force results from two-pion exchanges, such as those
in Fig.(4d).  The latter processes had been previously worked out by Coon and
Friar\cite{2} using a chiral Lagrangian in tree order, but without considering
power counting or the short-range interactions.  These results appeared to
differ but were later shown to be merely different off-shell extensions of
OPEP\cite{19}, which is not unique, as we remarked earlier. Nevertheless, a
mapping exists that exactly transforms one result into the other in leading 
order.

The implicit form of OPEP used by Weinberg is an energy-dependent potential,
which in most many-body computational procedures is intractable.  Precisely the
same problem had arisen long ago in the two-nucleon problem, with competing BW
(Brueckner-Watson\cite{20}) and TMO (Taketani-Machida-Ohnuma\cite{21})
potentials.  The former, which corresponds to the Weinberg $3N$ force, is
energy-dependent (although BW ignored those terms), while the latter is
energy-independent and corresponds to the Coon-Friar $3N$ force.  Leading-order
chiral two-pion-exchange $NN$ forces were first calculated by Ord\'o\~nez, Ray,
and van Kolck\cite{22} and subsequently verified by Friar and Coon\cite{19}. 
Although this is a good start toward a nucleon-nucleon potential with proper
chiral constraints, it remains to be seen whether $NN$ scattering is sensitive
to these nuances. This is being examined in Nijmegen. 

Of what use is power counting?  A generic Lagrangian for pions $(\vec{\pi})$,
nucleons ($\psi$), and photons ($A^{\mu}$), and containing derivatives,
$\partial^{\mu}$, can be pieced together with the scales we introduced earlier
to produce a series of terms, each with the form 
$$
L \sim c_{l m n} 
\biggl[ \frac{\overline{\psi}\psi}{f^2_{\pi} \Lambda} \biggr]^l
\biggl[ \frac{\vec{\pi}}{f_{\pi}} \biggr]^{m} 
\biggl[ \frac{\partial^{\mu}, A^{\mu}, m_{\pi}}{\Lambda} \biggr]^n 
f^2_{\pi} \, \Lambda^2 \, . \eqno (5.1)
$$
This can be motivated by looking at the form of the pion and nucleon masses and
free energies.  The dimensionless coefficients should be of order (1) if naive
dimensional power counting\cite{23} holds, which leads to a ``natural'' theory. 
Of utmost importance is the chiral constraint\cite{18}, which is conventionally
written in terms of the number of derivatives and nucleon fields at each vertex
$$
\Delta = l + n - 2 \; \geq \rm \; 0 \, , \eqno (5.2)
$$
which guarantees that no $\Lambda$ occurs in the numerator of Eq.(5.1). If the
power series implied in Eq.(5.1) converges for nuclei and if the various $c$'s
are of order (1), then nuclei are ``soft'' and ``natural''. 

Several examples appropriate to nuclei illustrate these ideas.  Constructing a
Walecka-type model\cite{17} with a scalar isoscalar channel and a vector
isoscalar channel in zero-range form(but with no pions), one finds from
Eq.(5.1)
$$
L_W = \alpha_s \frac{(\overline{\psi} \psi)^2}{f^2_{\pi}} +  
\alpha_v \frac{(\overline{\psi} \gamma^{\mu} \psi)^2}{f^2_{\pi}} 
\, , \eqno (5.3)
$$
with values of $\alpha_s = -1.98$ and $\alpha_v = 1.48$ obtained from a recent
Dirac-Hartree calculation\cite{24}.  Thus these coefficients are natural and,
in fact, are quite typical. 

The second example concerns the convergence of the series implied in Eq.(5.1) as
a function of the nuclear density, $\rho$.  Since $\overline{\psi} \psi \sim
\rho$, and since the density of nuclear matter, $\rho_{\rm n m} 
\sim 1.5 f^3_{\pi}$, we have 
$$
L \sim [c_l \sim 1] [1.5 f^3_{\pi} / f^2_{\pi} \Lambda]^l 
\sim \biggl[ \frac{1}{7} \biggr]^l \, , \eqno (5.4)
$$
which is fairly rapid convergence.  Unfortunately, the number of terms in this
expansion grows explosively as $l$ increases.  At higher densities convergence
will be worse. 

The third example concerns the ancient art of potential fabrication. In the old
days it was found that naive PS coupling of pions and nucleons($\gamma_5$) led
to very strong forces at every order.  The problem was that PS coupling
optimally connects nucleon and antinucleon spinors, leading to very large and
unphysical ``pair'' contributions to nuclear forces. An \underline{ad}
\underline{hoc} procedure was invoked called ``pair suppression'', which deleted
such terms.  It is easy to show\cite{2} that these unphysical pair terms
correspond to a model with $\Delta = -1$, and thus Eq.(5.2), which forbids this
value, is equivalent to ``pair suppression'' in nuclear physics. It is also easy
to show that $\Delta < 0$ leads to very strong many-body forces, which would
make nuclear physics calculationally intractable. 

\section{The Power of Counting}

One typically counts powers\cite{25} of small momenta in an amplitude:
$\overline{p}^{\nu}$. This procedure is quite old and is very useful, but it
fails to describe the nuclear problem unless modifications are made. A nucleus
is a self-bound system that shares its available momentum; if one nucleon draws
an amount from the ``bank'', less is available for the others. It is this
mechanism that weakens many-nucleon forces.  Weinberg's final power-counting
rules\cite{18} take this into account. A simplified version is illustrated
below.  Ignoring isospin factors and other factors $\sim 1$, the
one-pion-exchange potential has the form 
$$
V_{\pi}(\mbox{\boldmath$r$}) \sim \biggl[ \frac{1}{f^2_{\pi}} \biggr]
\int \frac{d^3 q}{(2 \pi)^{3}} \biggl[ \frac{\mbox{\boldmath$\sigma$}(1)  
\cdot \mbox{\boldmath$q$} \mbox{\boldmath$\sigma$}(2) \cdot  
\mbox{\boldmath$q$}}{\mbox{\boldmath$q$}^2 + m^2_{\pi}} \biggr]
e^{i \mbox{\boldmath$q$} \cdot \mbox{\boldmath$r$}} . \eqno (6.1)
$$
The basic amplitude for OPE (in brackets) has dimension $\nu = 0$.  The phase
space factor has an additional dimension, $\Delta \nu = 3$.  We add the two
together and count this operator as having dimension $\nu = 3$.  Three-nucleon
operators have an additional dimension $\Delta \nu = 6$, etc. Failure to
implement the momentum sharing makes it impossible to compare operators
involving different numbers of nucleons for fixed A (i.e., in a given nucleus). 

Finally, we can write a very simple and elegant power-counting formula for the 
nuclear (potential) case
$$
\nu = 1 + 2 (n_c + L) + \Delta \, , \eqno (6.2)
$$
where $L$ is the number of loops, $\Delta$ is the sum of the individual
Lagrangian power-counting factors ($l + n -2 \geq 0$), and $n_c$ is a
topological factor:  the number of nucleons interacting with {\underline{at}
\underline{least} one other minus the number of clusters with \underline{at}
\underline{least} two nucleons interacting.  A constant that depends only on the
total number of nucleons in the nucleus has been dropped.  In the most common
configuration a single cluster of $N$ nucleons interacts, which leads to $n_c =
N - 1$, while in lowest order one has $L = 0$ and $\Delta = 0$. This is the
leading-order $N$-body-force case, and we find 
$$
V_{Nbf} \sim \left(\frac{\overline{p}}{\Lambda} \right)^{2N-1}\, . \eqno (6.3)
$$
Thus, $N$-body forces weaken \underline{progressively} and geometrically because
of chiral symmetry $(\Delta \geq 0)$. This result has enormous implications for
nuclear physics. 

\section{Quantitative Tests and Estimates}

We will examine three areas of few-nucleon physics:  (1) charge-independent
nuclear forces; (2) isospin-violating nuclear forces; (3) meson-exchange
currents. 

Dimensional power counting suggests that OPEP and the short-range nuclear forces
should be comparable (both have $\nu = 3$), but they are not.  The reason is 
simple. Strong short-range repulsion produces a coherent effect (a barrier, 
whose penetration is suppressed). This in turn produces a hole in the wave 
function which always ``wins'', no matter how strong the potential, and OPEP 
therefore dominates. Two-pion-exchange(TPE) is predicted to be suppressed 
because it corresponds to $\nu = 5$.  Although most realistic potentials have a
phenomenological component of two-pion range, there is not yet any direct
quantitative evidence that those components play a significant role in $NN$
scattering (note that we are not discussing $\rho$-meson exchange, but rather
the long-range tail of the force in the TPE channel). Tests of the TPE potential
are planned at Nijmegen. 

The most important evidence for chiral symmetry in nuclei is that nuclear
physics is tractable.  The recent\cite{7} calculation of $A = 2-6$ used a
realistic $NN$ force model and a weak $3N$ force adjusted to fit $^3$H; it also
fits $^4$He.  Their results are in good accord with power-counting predictions
$$
\langle V_{NN} \rangle \sim 20\, {\rm MeV/pair} \, , \eqno (7.1a)
$$
$$
\langle V_{3Nf} \rangle \sim 1\, {\rm MeV/triplet} \, , \eqno (7.1b)
$$
$$
\langle V_{4Nf} \rangle \lsim \, 0.1\, {\rm MeV/quartet} \, , \eqno (7.1c)
$$
since (leading-order) $NN$ forces correspond to $\nu = 3$, $3N$ forces to 
$\nu = 5$ and $4N$ forces to $\nu = 7$.  The latter are therefore likely to be 
negligible. 

Finally, we note that relativity appears to be a correction in few-nucleon
systems.  While not specifically a consequence of chiral symmetry, an
expansion in powers of $1 / M \sim 1 / \Lambda$ is intimately related to the
usually power counting.  We note that regularizing (in the nuclear physicist's
fashion) at a momentum $\sim \Lambda$ generates relativistic corrections $\sim$
5\% to nuclear energies. This result is somewhat controversial and requires more
study (particularly using fully-relativistic calculations), but is unlikely to
be qualitatively wrong. 

Isospin violation in the nuclear force is still a rather poorly understood
phenomenon.  The isospin dependence of the nuclear force is classified according
to four categories\cite{26}, with two-nucleon operators having forms 
\begin{tabbing}
xxxxx\=xxxxxxxxxxxxxxxxxxxxxxxxxxxxxxxx\=xxxxxxxxxxxxxxxxxxxxxxxxxx\= \kill
{} \> {(I).  1 and $\mbox{\boldmath$t$}_1 \cdot \mbox{\boldmath$t$}_2$} \>  
{charge independent (CI)} \\
\\
\> (II).  $t^z_1 t^z_2 - \mbox{\boldmath$t$}_1 \cdot \mbox{\boldmath$t$}_2/3$  
\> charge-independence breaking (CIB) \\
\\
\> (III).  $(t_1 + t_2)_z$ \> charge-symmetry breaking (CSB) \\
\\
\> (IV).  $(t_1 - t_2)_z$ and $(\mbox{\boldmath$t$}_1 \times  
\mbox{\boldmath$t$}_2)_z$ \> charge-symmetry breaking (np only) \\  

\end{tabbing}     
where $\mbox{\boldmath$t$}_1$ and $\mbox{\boldmath$t$}_2$ are the isospin
operators of nucleons ``1'' and ``2''. The class (II) operator is an isotensor,
while (III) and (IV) are isovectors.  The class (IV) operators are 
nonvanishing only for the $np$ system, while class (III) vanishes for the $np$ 
system. 

It was shown by van Kolck\cite{25} that I $>$ II $>$ III $>$ IV.  He did this by
constructing effective chiral Lagrangians corresponding to $d - u$ quark mass
differences, characterized by $\epsilon = \frac{m_d - m_u}{m_d + m_u} \sim 0.3$,
which has the isospin character of $t_z$.  A separate construction of effective
Lagrangians resulting from freezing out hard-photon exchanges is characterized
by the fine-structure constant, $\alpha$. These two Lagrangian ``ladders'' (each
corresponding to different powers of small momenta) must be realigned so that
the separate dimensionless factors of $\epsilon$ and $\alpha$ are taken into
account.  Using the mass shifts of neutral and charged mesons and nucleons as
input, van Kolck argued that quark-mass terms were dominant in leading order. 
The relevant parts of his Lagrangian are
$$
L \sim - \frac{g_A}{f_{\pi}}\, \overline{N} \mbox{\boldmath$\sigma$} \cdot 
\mbox{\boldmath$\nabla$} \mbox{\boldmath$t$} \cdot \mbox{\boldmath$\pi$} N 
+ {\rm [mass-splitting \; \;terms]} + \frac{\beta_1}{2 f_{\pi}} \, 
\overline{N} \mbox{\boldmath$\sigma$} \cdot \mbox{\boldmath$\nabla$} \pi_0 N
$$
$$
+ \gamma_s \overline{N} \; t_3 N \; \overline{N} N \; + \gamma_{\sigma}
\overline{N} \mbox{\boldmath$\sigma$} t_3 N \cdot \overline{N} 
\mbox{\boldmath$\sigma$} N  \, . \eqno (7.2)
$$
The first term is the usual $\pi NN$ coupling that conserves isospin, while the
remaining terms violate isospin.  The very important mass-splitting interactions
for charged and neutral pions and nucleons are next, followed by an
isospin-violating $\pi^0 NN$ coupling and two short-range $NN$ terms.  Nuclear
physicists have approached calculations of isospin-violating mechanisms in terms
of resonance-saturation diagrams of the type shown in Fig.(5a).
Pseudoscalar-meson exchange\cite{27} can incur mixing (at the blob) that
converts a $\pi^0$ to an $\eta$ (or $\eta^{\prime}$).  In the limit of large
$\eta$ mass this produces $\beta_1$, which should be on the order of $(\epsilon
m ^2_\pi/\Lambda^2)$. Numerical evaluation of the mixing diagram using typical
constants leads to $\beta_1 \sim \frac{2}{3} (\epsilon m^2_{\pi}/\Lambda^2)$,
which is a reasonable result.  The same procedure can be applied to vector-meson
exchange\cite{27} and leads to the $\rho - \omega$ mixing depicted in Fig.(5b)
and to $\gamma_s$, which should be of order $(\epsilon m^2_{\pi}/f^2_\pi
\Lambda^2)$. Evaluating $\gamma_s$ in terms of the $\rho - \omega$ mixing model
with typically used parameters\cite{27,28} one finds $\gamma_s \sim \frac{1}{2}
(\epsilon m^2_{\pi}/f^2_\pi \Lambda^2)$, a not unreasonable value. There have
been many recent claims that the $\rho - \omega$ mixing parameter (fitted
on-shell for $q^2 \sim m^2_{\rho}$) has a strong off-shell suppression when
taken into the nuclear regime $(q^2 \leq 0)$. Various calculations ranging from
QCD sum rules to quark models have advanced this claim\cite{29}. Chiral
Perturbation Theory, on the other hand, makes no statements about specific
models.

\begin{figure}[htb]
  \epsfig{file=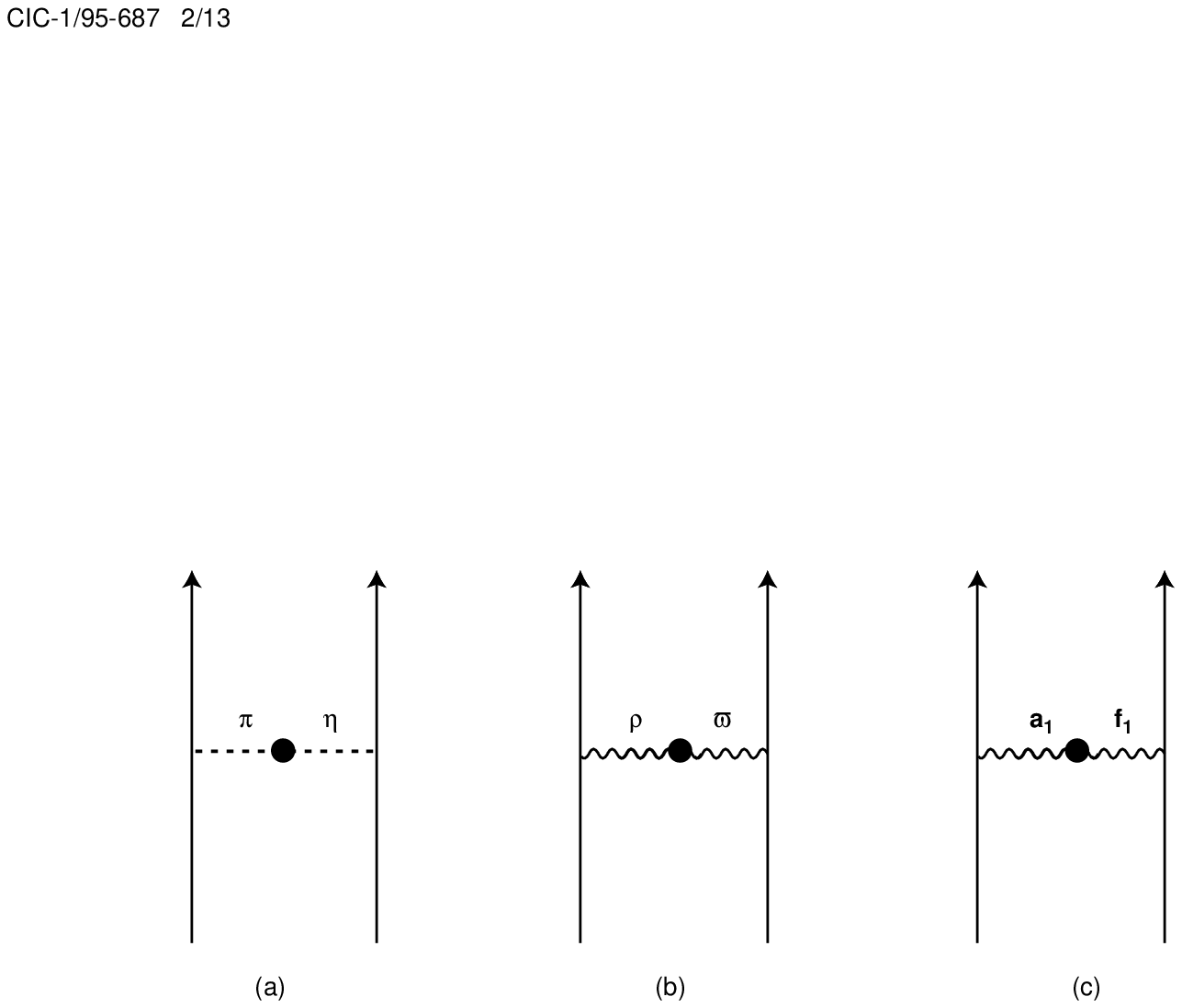,height=2.0in,bbllx=5pt,bblly=435pt,bburx=450pt,bbury=600pt,clip=}
  \caption{Meson-mixing processes that contribute to the isospin-violating
$NN$ force, including pseudoscalar-meson exchange in (a), vector-meson exchange
in (b), and axial-vector-meson exchange in (c).}
\end{figure}

Finally, pseudovector-meson exchange can take place (Fig.(5c)) and leads to a
spin-spin force in leading order.  Close-lying isospin doublets such as ($a_1,
f_1$) could contribute substantially to the $\gamma_{\sigma}$-term.  To the best
of our knowledge no such mechanism has ever been proposed, but it arises quite
naturally. 

Given a variety of isospin-violating mechanisms (which are all $\lsim$ 1\% of
the isospin-conserving force), which of them are the largest and how important
are they in nuclear physics?  Because OPEP is dominant the pion-mass-difference
term is very large ($\sim$ 3\% of OPEP) and leads to a large class II force.
Charge-symmetry breaking of class III is next in size and is of both conceptual
and practical importance in the few-nucleon problem.  Class IV forces require
antisymmetric complementary spin-space operators and are therefore the smallest.

Class III forces are a very important aspect of the $NN$ interaction and are
part of one of the most important success stories in few-nucleon physics.  The
$^3$He - $^3$H mass difference can be written as the difference of the
individual nucleon masses plus a (positive) binding energy difference, with
$^3$He being less bound than $^3$H.  Because $^3$He has a $pp$ pair and $^3$H an
$nn$ pair (each having two $np$ pairs), we expect that most of the 764 keV
binding energy difference is due to the Coulomb interaction between the two
protons in $^3$He, which generates 648(4) keV.  Small contributions from the
repulsive magnetic moment interactions, the motion of the protons, the $n-p$
mass difference in the kinetic energy, and similar small mechanisms generate
45(5) keV.  The remaining contribution has a short-range nature.  After removal
of electromagnetic effects from their interactions the $T = 1$ s-wave scattering
lengths of nucleons ($^1{\rm S}_0$ state) have the values\cite{28}: 
$np$(-23.7 fm), $pp$(-17.3(4) fm), $nn$(-18.8(3) fm).  The difference of $nn$ 
and $pp$ is -1.5(5) fm, which generates approximately 66(22) keV binding energy
difference\cite{7}. The total\cite{30} of 759(25) keV agrees well with
experiment:  764 keV. This impressive success in few-nucleon physics does,
however, raise one significant question.  If $\rho - \omega$ mixing was greatly
overestimated in the past, are there enough other mechanisms of sufficient size
to account for the -1.5(5) fm difference in scattering lengths? 

In recent years the Nijmegen PSA has successfully measured the $\pi N N$
coupling constants, of which there are three, by focusing on the long-range part
of the nuclear force\cite{14}.  Defining 
$$
f^2 = \left. \left(\frac{g_A m_{\pi^+} \, d}{2 f_{\pi}} \right)^2 \right/ 
4 \pi \, , \eqno (7.3)
$$
where $d - 1$ is the Goldberger-Treiman discrepancy\cite{31}, one can transform
measurements of $f^2$ into measurements of $d$.  The exchange of a neutral pion
between a pair of protons generates $f^2_{\pi^{0}pp} =$ 0.0751(6), while a
neutral pion exchanged between an $np$ pair gives $f_{\pi^{0}nn} f_{\pi^{0}pp}
=$ 0.0752(8).  Charged-pion exchange in the latter case gives $f^2_{\pi^{c}np}
=$ 0.0741(5).  These three pieces of experimental information can be analyzed in
terms of three other quantities:  the isospin-symmetric $d$; the quantity
$\beta_1$; a consistency condition, c, which should equal 1.  One finds\cite{32}
$$
d - 1 = 2.0(5){\rm \%} \, ,\eqno (7.4a)
$$
$$
\beta_1 = 1(8) \cdot 10^{-3} \, , \eqno (7.4b)
$$
$$
c = 1.007(6) \, . \eqno (7.4c)
$$
The value of $d - 1$ is considerably smaller than older values and corresponds
to a (monopole) form factor mass of $\sim$1 GeV.  Dimensional estimates of
$\beta_1$ are $\sim 6 \cdot 10^{-3}$, consistent with Eq.(7.4b); Eq.(7.4c) is
also satisfactory.  Because a huge amount of data was analyzed to obtain these
results, it will be extremely difficult to do better. 

Finally, we complete the circle with a discussion of meson-exchange currents. As
discussed thoroughly by Park, Min, and Rho\cite{8}, it is impossible to
understand the \underline{isovector} magnetic form factors in nuclei without
incorporating the effect of pion degrees of freedom.  The same observation holds
for the time component of the nuclear axial current\cite{33}. A number of 
processes contributing to nuclear electromagnetic interactions are shown in 
Fig.(6).

\begin{figure}[htb]
  \epsfig{file=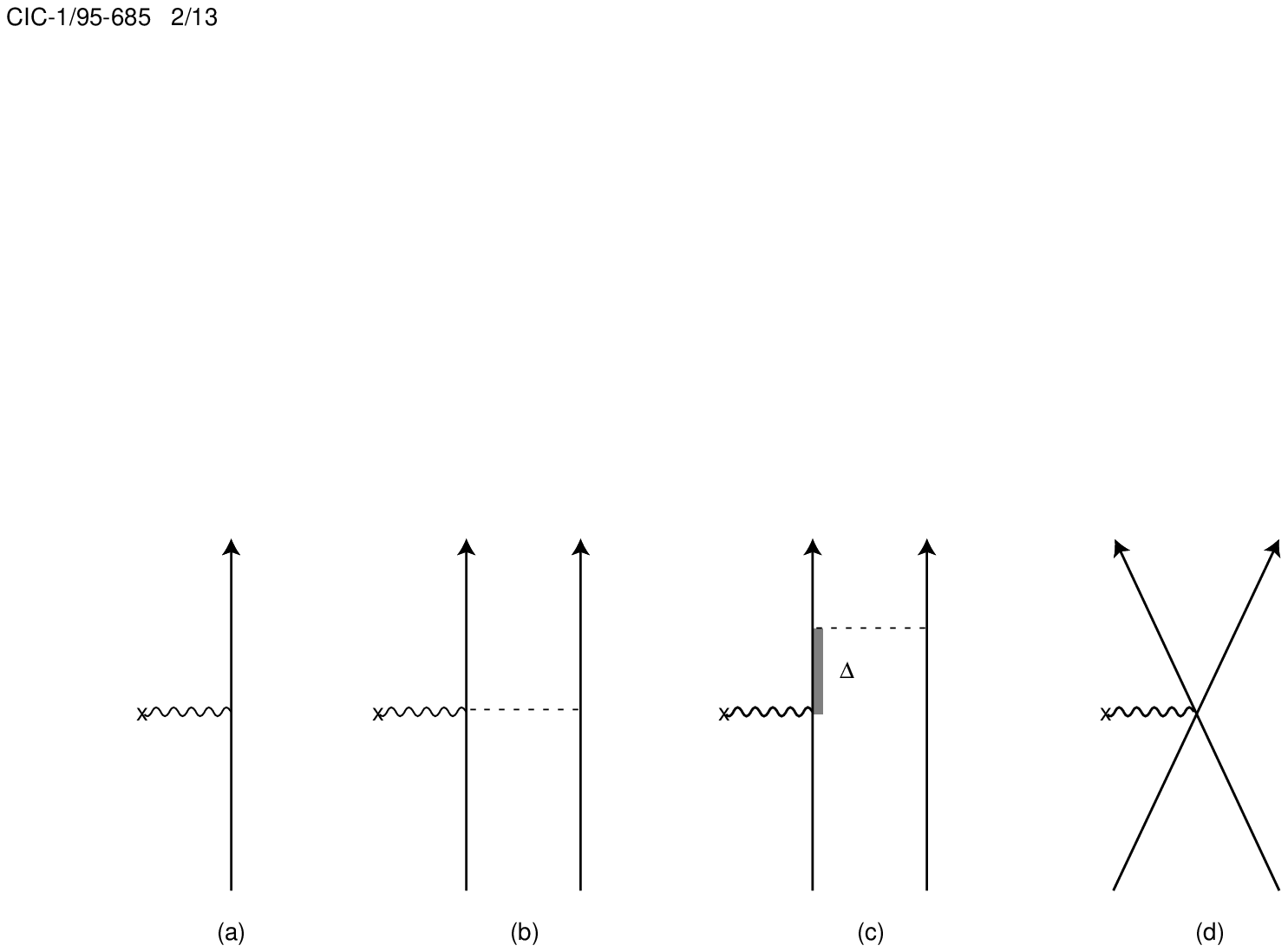,height=2.0in,bbllx=41pt,bblly=435pt,bburx=483pt,bbury=600pt,clip=}
  \caption{Nuclear electromagnetic interactions, with the impulse
approximation shown in (a), seagull MEC depicted in (b), isobar-mediated MEC
illustrated in (c), and short-range MEC sketched in (d).  Solid lines are
nucleons, dashed lines are pions, and wavy lines are (virtual) photons.}
\end{figure}

What about the effect of short-range (e.g., $\rho$) meson exchange?  This  
question was answered by Rho\cite{12}, who noted that power counting for meson-%
current operators has the form
$$
\nu = 1 + 2 (n_c + L) + \Delta + \delta \, , \eqno (7.5)
$$
where $\Delta \geq -1$ for an electromagnetic vertex (but $\geq 0$ for a strong
vertex) and $\delta$ is a spinor-reduction factor that equals 0 (for ``even''
operators) or 1 (for ``odd'' operators).  Specializing to the space part of the
vector current, we find that the impulse approximation $(\sim \mbox{\boldmath
$\gamma$}$, corresponding to $\delta = 1$) has $\nu = 1$ in leading order and is
shown in Fig.(6a).  The usual seagull MEC in Fig.(6b) has $\nu = 2$, and the
pion-exchange diagrams are therefore the largest corrections to the impulse
approximation.  Heavy MEC (Fig.(6d)) and isobar-mediated pion-range MEC
(Fig.(6c)) have $\nu = 4$.  The suppression of heavy-meson exchange arises
because insertion of a photon into a meson propagator involves cutting that
propagator, thus making two of them.  The extra propagator $\sim 1 / m^2_{\rho}
\sim 1 / \Lambda^2$, and this suppresses Fig.(6d) by two powers of $\nu$
compared to Fig.(6b).  This is a lovely result! 

Park, Min, and Rho\cite{8} calculate all contributions to the (magnetic dipole)
$np$ radiative capture transition with $\nu \leq 4$ and obtain excellent
agreement with experiment.  In view of the complexity of the calculation and its
close relationship to chiral perturbation theory calculations in the one-nucleon
sector, this is an important technical achievement. 

\section{Conclusions}

One-pion exchange dominates in the binding of light nuclei and in meson-exchange
currents. This follows from power counting and the coherent suppression
resulting from barrier penetration at short distances.  Chiral symmetry provides
order in nuclear forces:  without this symmetry nuclear physics would be
intractable. Turning the argument around, the tractability of nuclear physics
provides strong evidence for chiral symmetry, which weakens $N$-body forces as
$N$ increases and $n$-pion exchanges compared to OPEP.  The sizes of various
meson-exchange currents can be understood in terms of dimensional power 
counting. Mechanisms for isospin violation in the nuclear force are also
consistent with dimensional power counting. Finally, few-nucleon systems
continue to be the testing ground for new ideas in nuclear physics because of
our ability to calculate accurately in those systems.

\section{Acknowledgements}

This work was performed under the auspices of the U.\ S.\ Department of Energy. 
Numerous discussions with J.\ de Swart were very helpful. I.\ Sick generously
contributed figures. I would like to dedicate this talk to Gerry Brown, who was
the first to argue that chiral symmetry was an essential ingredient of nuclear
forces\cite{34}, and to Bryan Lynn\cite{17} who has argued that CS provides a
framework for nuclear physics.

\end{document}